\documentclass[10pt,a4paper,onecolumn]{article}
\usepackage{amsmath,amssymb}
\usepackage{graphicx,subfigure}
\usepackage{dcolumn}
\usepackage{bm}

\newcommand{\beq}[1]{\begin{equation}\label{eq:#1}}
\newcommand{\eeq}{\end{equation}}

\begin{document}
\title{Shift of percolation thresholds for epidemic spread between static and dynamic small-world networks}
\author{
J. K. Ochab\footnote{jeremi.ochab@uj.edu.pl}
\and
P. F. G\'ora\footnote{Also at Mark Kac Complex Systems Research Centre.}\\
M. Smoluchowski Institute of Physics, Jagellonian University\\ Reymonta 4, 30-059 Kraków, Poland
}

\maketitle
\begin{abstract}
The aim of the study was to compare the epidemic spread on static and dynamic small-world networks. The network was constructed as a 2-dimensional Watts-Strogatz model ($500\times500$ square lattice with additional shortcuts), and the dynamics involved rewiring shortcuts in every time step of the epidemic spread. The model of the epidemic is SIR with latency time of 3 time steps. The behaviour of the epidemic was checked over the range of shortcut probability per underlying bond $\phi=0-0.5$. The quantity of interest was percolation threshold for the epidemic spread, for which numerical results were checked against an approximate analytical model. We find a significant lowering of percolation thresholds for the dynamic network in the parameter range given. The result shows that the behaviour of the epidemic on dynamic network is that of a static small world with the number of shortcuts increased by $20.7\pm1.4 \%$, while the overall qualitative behaviour stays the same. We derive corrections to the analytical model which account for the effect. For both dynamic and static small-world we observe suppression of the average epidemic size dependence on network size in comparison with finite-size scaling known for regular lattice. We also study the effect of dynamics for several rewiring rates relative to latency time of the disease.
\end{abstract}

\section{\label{sec:intro}Introduction}
The epidemic modelling has become a significant and needed branch of complex systems research, as we have witnessed the recent epidemic threats and outbreaks of human diseases (H5N1 and H1N1 influenzas~\cite{WHOswine,WHOavian} or severe acute respiratory syndrome~\cite{WHOsars,Dye}) or animal (foot-and-mouth disease~\cite{Keeling}) and plant diseases alike (e.g. Dutch elm disease~\cite{Gilligan} or rhizomania~\cite{Stacey}). There are two crucial characteristics of the epidemic spread that make it complicated to be modelled on the one hand, and costly to be prevented in reality on the other: firstly, a number of infectious diseases exhibit long-range transmissions of varied nature, and secondly, the contact network of individuals affected by the disease may change in time as the epidemic spreads. These features make epidemiological models a part of larger studies of dynamics \textit{on} complex networks, but also dynamics \textit{of} complex networks.

Research findings of the epidemic spread on dynamic networks include its behaviour on adaptive networks, where the susceptible are able to avoid contact with the infected~\cite{Blasius}), however a coupling between the epidemic and the network dynamics does not necessarily exist. For instance, in \cite{Dybiec}, spread of the aforementioned plant diseases is modelled by vectors performing random walk on the network, thus infecting individuals on their paths; Saram\"aki and Kaski~\cite{Kaski} utilise SIR (Susceptible-Infectious-Removed) mechanism on a dynamically changing small-world contact network, although mainly time development of the epidemic is of their interest. Likewise, in \cite{Meyers} (where focus is on the average epidemic size in time) nodes of the contact network can swap their edges at a given rate, preserving the degree distribution. It is also worth to note \cite{Eames}, where disease spread was simulated on a weighted contact network produced from \textit{real} day-to-day encounters (as weights represent the frequency of encounters, the dynamics has been in a sense projected onto static weighted network).

While dynamic network models have been applied in the recent research, it seems that we lack comparative study on how the dynamics of the network influences the process that takes place on it. The aim of this paper is to find and quantify this effect for SIR epidemic spread on static and dynamic small-world networks. Based on known analytical calculations for static small-world network~\cite{Newman} we derive corrections accounting for the dynamics of the network, and check the results against numerical agent-based simulations.

\section{\label{sec:model} Model}
  \subsection{\label{sec:net} Network }
We adopt Watts-Strogatz model of a small-world network \cite{Strogatz}: first we take a 2-dimensional square lattice with $N=L^2$ nodes and $2N$ undirected edges. To avoid some finite-size effects we impose periodic boundary conditions for the lattice (i.e. we get a torus). Then, we add a number of undirected edges between random pairs of nodes. The number of additional edges (`shortcuts') is set as $2 \phi N$, hence $\phi$ is shortcut probability per underlying bond. Network with $\phi=0$ is just a \textit{regular lattice}. For nonzero $\phi$ we call the network a \textit{static small-world}.

The third type of network is a \textit{dynamic small-world}. One can construct it by randomly distributing shortcuts in every time step of simulation. Here, we choose $2 \phi N$ nodes randomly, and keep them fixed for the whole run of the epidemic. In every time step we randomly launch shortcuts anchored in these nodes, which means the dynamics consists in rewiring one end of these shortcuts. For the sake of simplicity we allow for multiple shortcuts being incident with the same node, for shortcuts leading to nearest neighbours, and for loops being formed. The construction of the source nodes launching shortcuts allows for an easier interpretation of the network: the fixed nodes could correspond to centres of activity that can be identified as in the real world networks.

\begin{figure}[htbp]
\subfigure[]{
	\label{fig:1a}
	\includegraphics[width=0.30\linewidth]{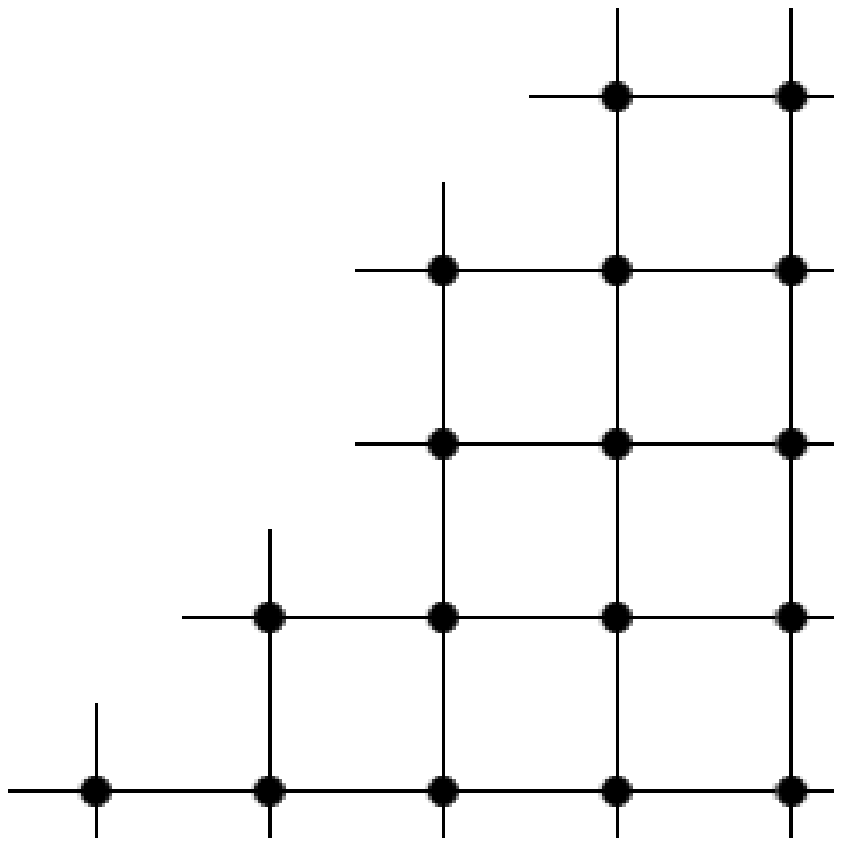}
}
\subfigure[]{
	\label{fig:1b}
	\includegraphics[width=0.30\linewidth]{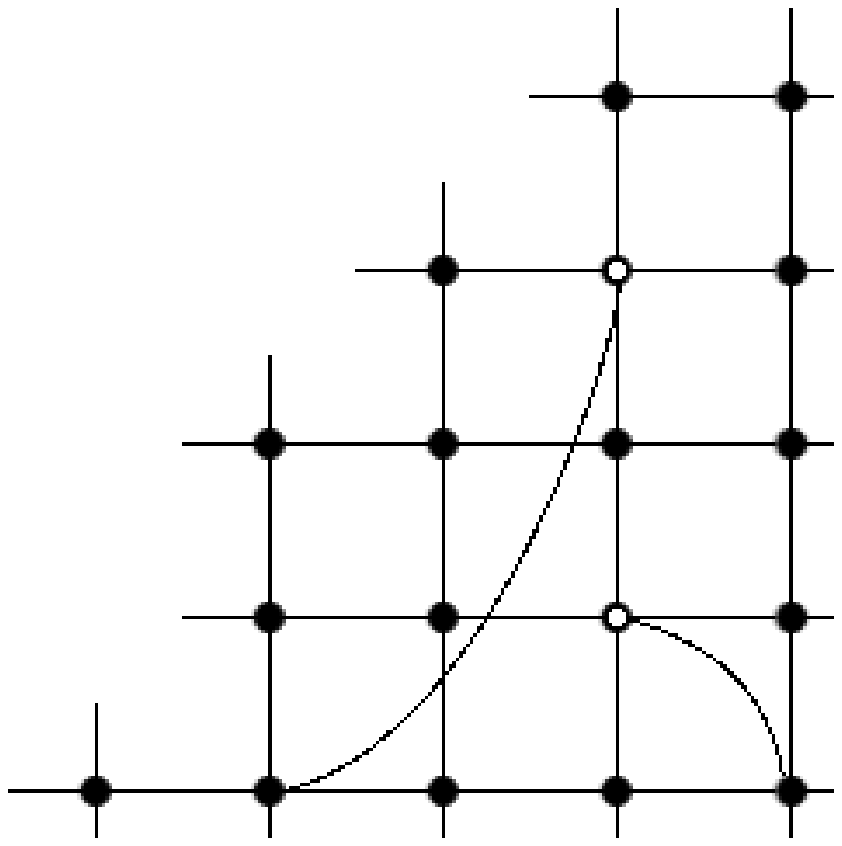}
}
\subfigure[]{
	\label{fig:1c}
	\includegraphics[width=0.30\linewidth]{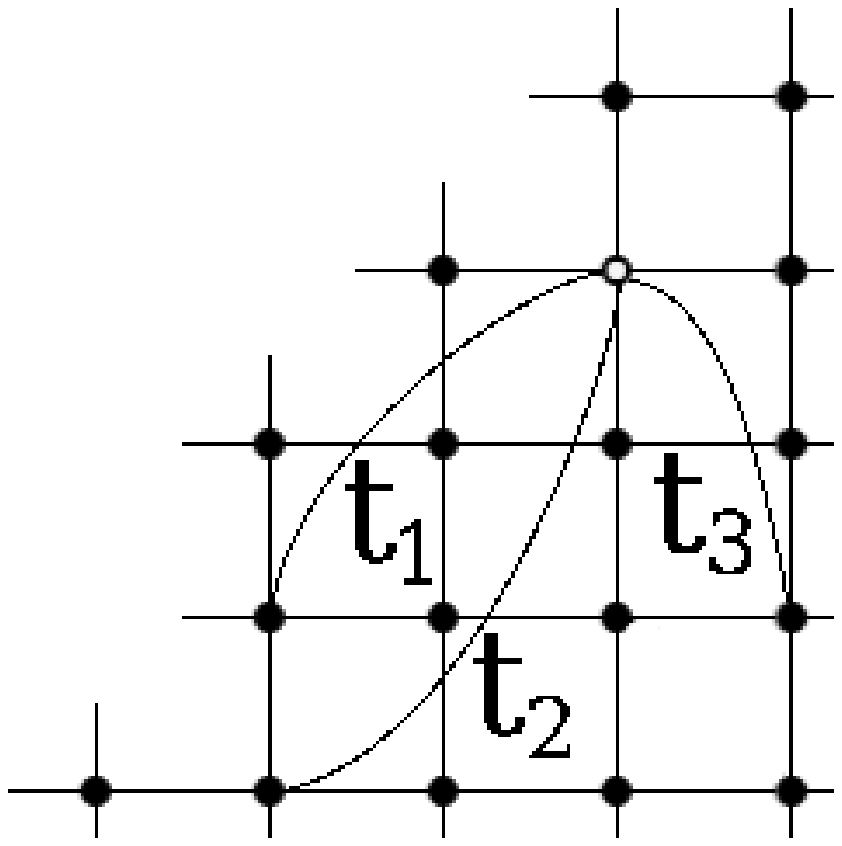}
}
	\caption{\label{fig:1} \subref{fig:1a} Regular 2D grid with periodic boundary conditions (torus). \subref{fig:1b} Watts-Strogatz 2D small-world network: 2D grid with shortcuts added to it. \subref{fig:1c} Dynamic small-world: all the long-range links connected to a set of source nodes randomly rewire in time.}
\end{figure}

  \subsection{\label{sec:epid} Epidemic }
\begin{figure}[htbp]
	\includegraphics[width=\linewidth]{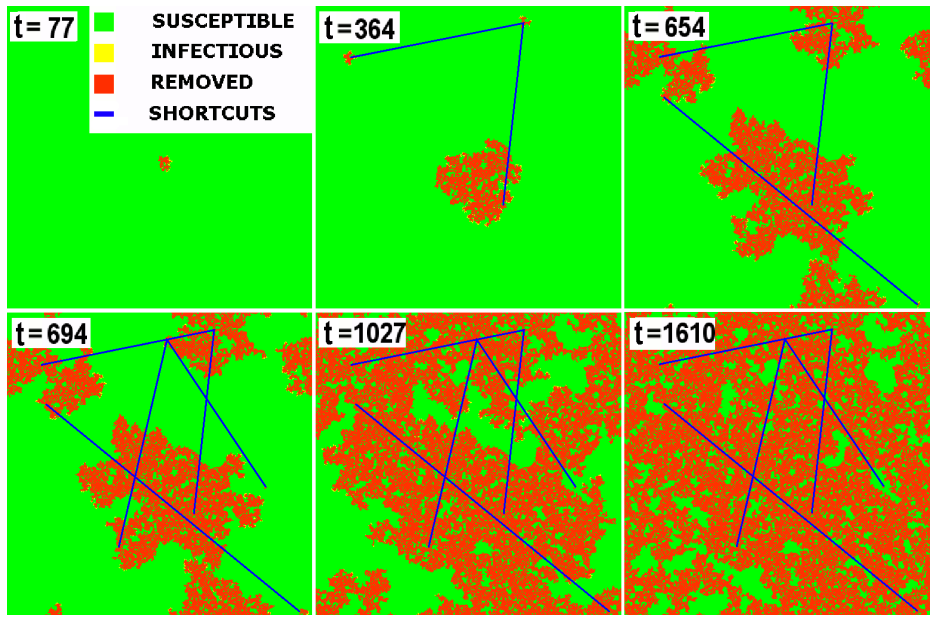}
	\caption{\label{fig:epid} (Colour online) Snapshots of the epidemic spread slightly above percolation threshold. $L=512$, the number of shortcuts is $10$ (which gives $\phi=2\cdot10^{-5}$). $t$ gives the epidemic's time steps. The snapshots for $t=364, 694$ show a dynamic infection (the two joined blue lines appear).}
\end{figure}

The SIR (Susceptible-Infectious-Removed) model is adopted, where the disease is transmitted along the edges of the network in discrete time steps. The probability $p$ of infecting a susceptible node by an infectious neighbour during one time step is set equal for short- and long-range links, both static and dynamic. The latency time $l$ of the disease is measured in discrete time units (we take $l=3,4$). Thus, an infectious node can transmit disease to susceptible nodes with probability $p$ every turn for the period of $l$ turns, and after that time it is removed, i.e. it cannot infect nor be reinfected. Every simulation starts with only one initially infecting node, all others being susceptible, and it ends when no node in the infectious state is left. Sample snapshots of the epidemic time development are presented in Fig.\ref{fig:epid}.

Grassberger \cite{Grass} related the probability of infection to the probability $T$ in bond percolation through $T=\sum_{t=1}^{l} p(1-p)^{t-1}=1-(1-p)^{l}$, where $T$ is the so called \textit{transmissibility} (it is the total probability of a node infecting one of its neighbours during the whole latency time). In the case of 2-dimensional square lattice the bond percolation threshold is $T_c=0.5$.

\section{\label{sec:numerical} Numerical data}
  \subsection{\label{sec:params} Parameters of simulations }
The linear lattice size used for most calculations is $L=\sqrt{N}=500$. In Section \ref{sec:results_scaling} we take sizes $L=50, 63, 79, 100, 126, 158, 199, 251, 315, 397, 500$. The disease latency is set to $l=3$ (for faster simulations reported in Sec.~\ref{sec:results}) or  $l=4$ (in Sec.~\ref{sec:dynamics} in order to get larger set of dynamic rates). The range of probability $p$ scanned is $p=0.05-0.22$ (depending on $\phi$) with resolution of $1/1024$, which translates to around $T=0.15-0.5$. For every $p$ and $\phi$ the epidemic is run $1024$ times with random distributions of shortcuts each time. The fraction of shortcuts is $\phi=0-0.5$, with steps of $0.025$. The simulations are performed for both static and dynamic small-world network.

  \subsection{\label{sec:threshold} Calculating percolation threshold }
In the study of the epidemic spread on networks, we stick to the percolation theory as a reference point. In the theory, a percolation threshold would be the value of $p$ that generates an epidemic cluster spanning between the boundaries of the whole system. Otherwise, it is possible to define percolation as the point at which a cluster of macroscopic size forms (i.e. it occupies a finite fraction of the system for $N\rightarrow\infty$). We employ the latter to define percolation threshold (numerically) as the point at which the average epidemic's size divided by $N$ rises above a certain value (here, set to $0.00115$). The average is taken over a number of reruns for different shortcut drawings. As we can perform simulations only for finite sizes, we take the results for a relatively large network of $\sqrt{N}=500$.

The choice of the threshold value is taken so as to calibrate the results for the static network to the previously confirmed analytical result. We take as the theoretical model \cite{Newman}, where the generating function and series expansion methods were used to find the approximate position of bond percolation transition in 2D small-world network, which corresponds to the epidemic spread on what we refer to as static small-world.

\section{\label{sec:anal} Theoretical analysis}


We can account for the change between static and dynamic networks analytically using the model known for static small-world network~\cite{Newman}. As the original theory has no time variable, it would be a hard task to introduce dynamics explicitly. The solution, however, is astonishingly simple. One can estimate the average number of nodes infected through shortcuts during latency time $l$: 
\beq{}
N_{stat}=\phi_{stat} N \cdot T= \phi_{stat} N \cdot \sum_{t=1}^{l} p(1-p)^{t-1} \quad,
\eeq
i.e. the number of shortcuts in the static network multiplied by the total probability of infecting a neighbouring node (this probability is the same for both regular links and shortcuts). The analogous expression for the dynamic network is found easily
\beq{teor}
N_{dyn}=\phi_{dyn} N \cdot \sum_{i=1}^{l} i \binom{l}{i} p^i (1-p)^{l-i} = \phi_{dyn} N \cdot l p \quad,
\eeq
where the sum is an average number of infections transmitted by a single source of dynamic shortcuts for a given latency time. It comes from the fact that a dynamic shortcut can pass infection several times (the factor $p^i$), while in the static case a node could infect only once (since nodes cannot be reinfected in the SIR model). This expression predicts lowering percolation thresholds, although numerical values of the shift are considerably smaller than the ones obtained from simulations.

\begin{figure}[htbp]
\subfigure[]{
	\label{fig:infekcje_a}
	\includegraphics[width=0.30\linewidth]{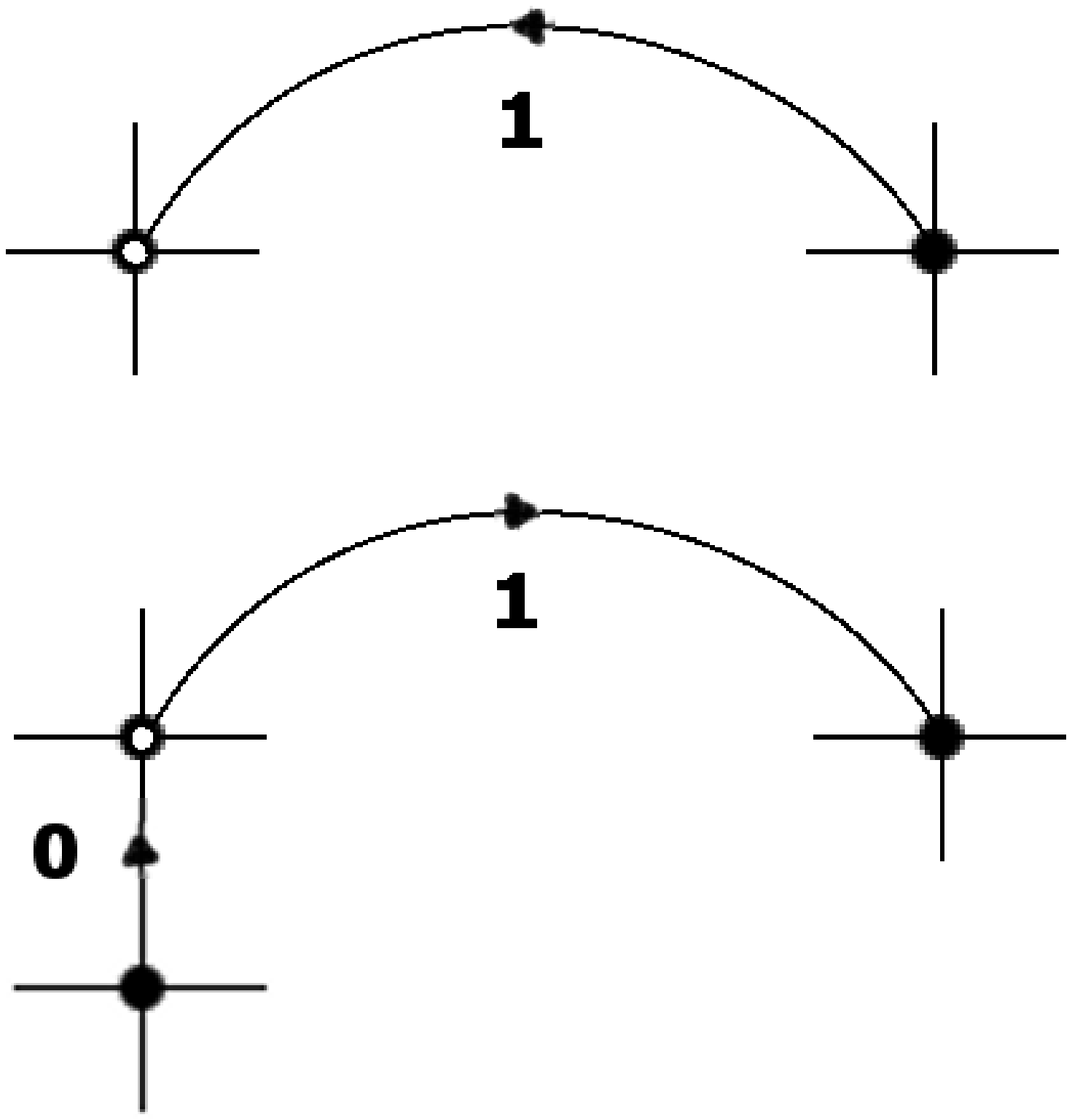}
	}
\subfigure[]{
	\label{fig:infekcje_b}
	\includegraphics[width=0.30\linewidth]{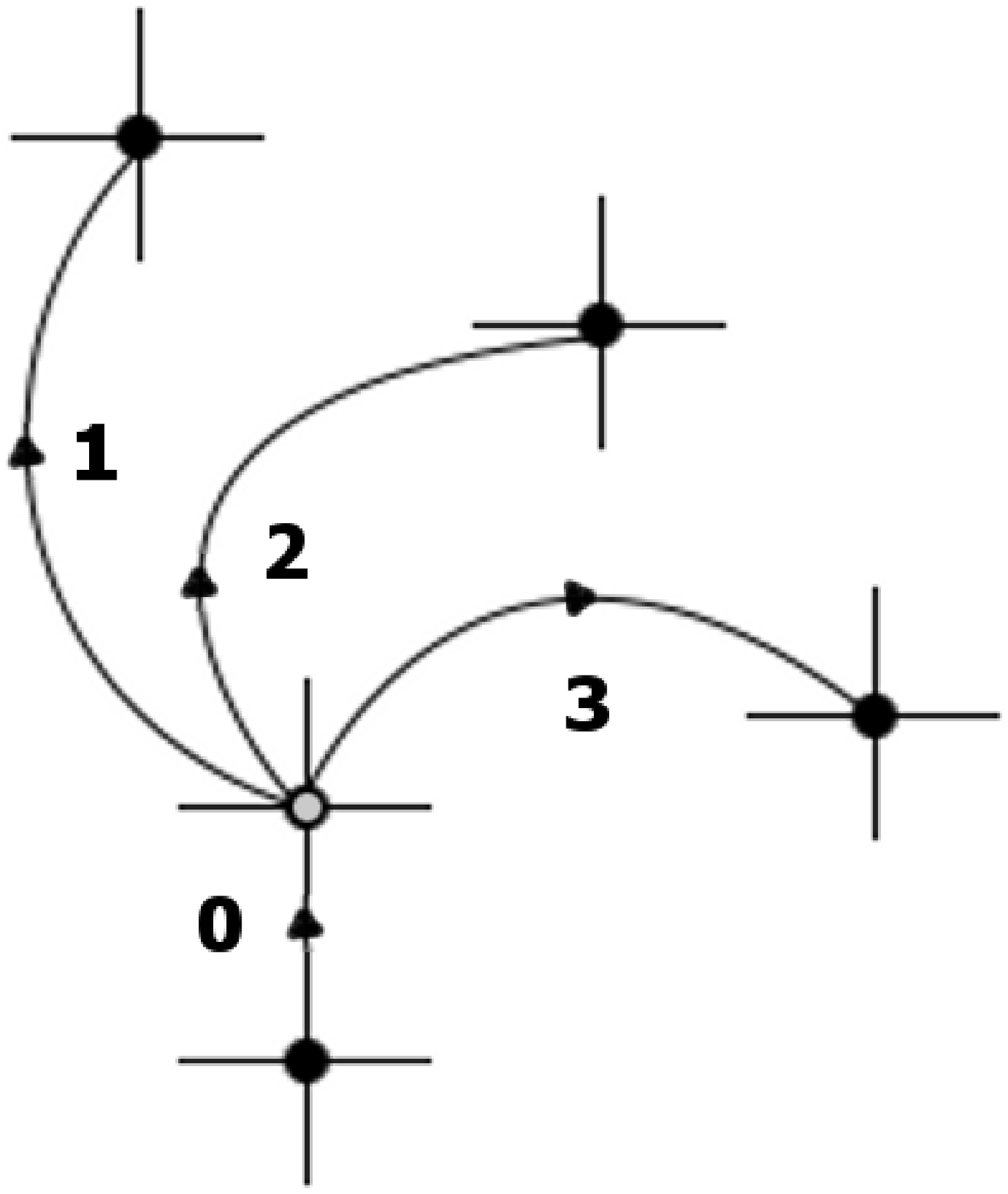}
	}
\subfigure[]{
	\label{fig:infekcje_c}
	\includegraphics[width=0.30\linewidth]{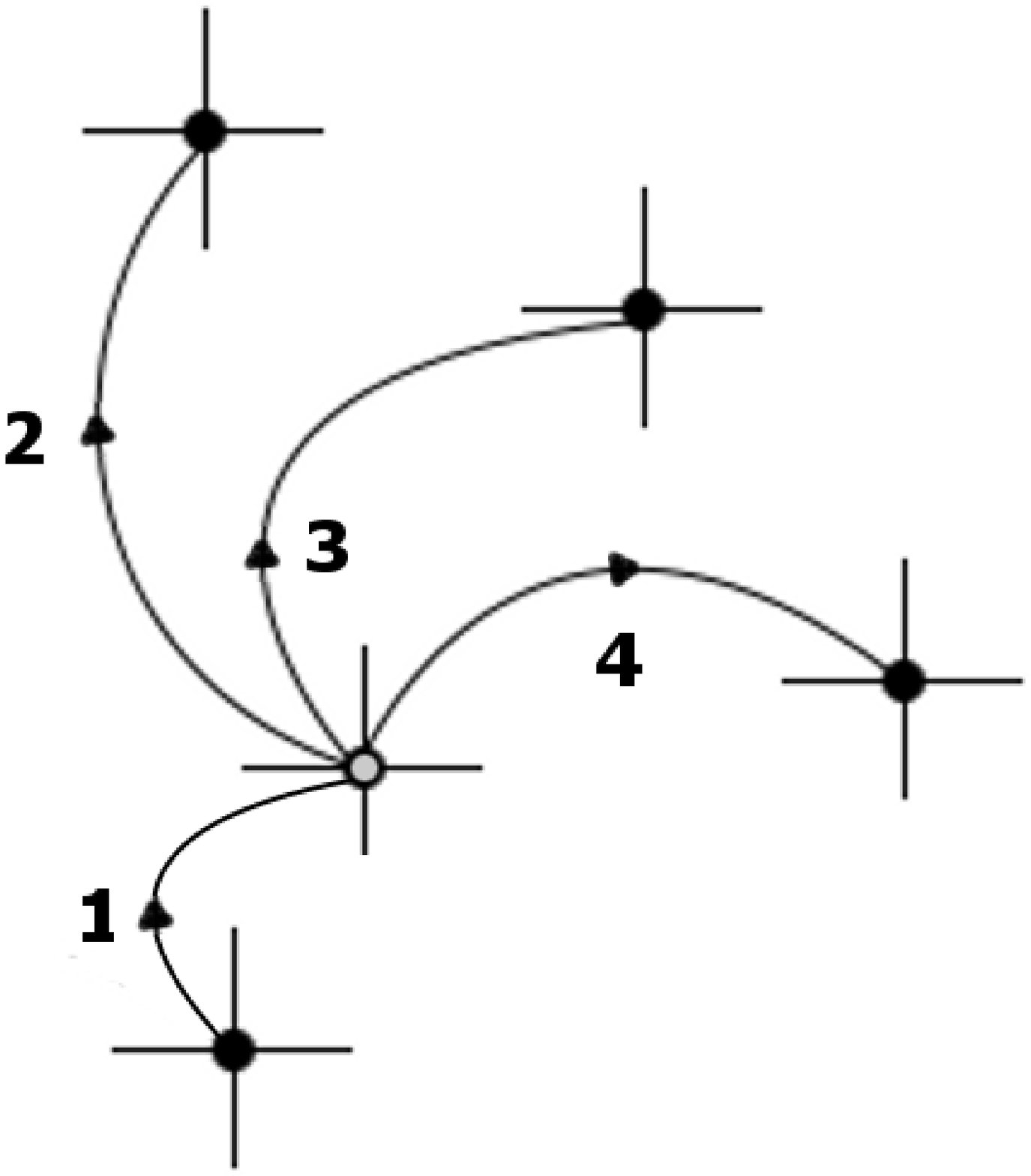}
	}
	\caption{	\label{fig:infekcje} \subref{fig:infekcje_a} Infections through static shortcuts are symmetric. \subref{fig:infekcje_b} Infection of the dynamic shortcuts' source through regular lattice. \subref{fig:infekcje_c} Infection of the dynamic shortcuts' source through a shortcut.}
\end{figure}

Figures \ref{fig:infekcje_a}-\ref{fig:infekcje_c} explain why the above expression is not yet correct: it is derived only for the source nodes passing the disease on, while it disregards the fact that the node may itself become infected via long-range link. Since on the static network there is no difference between shortcuts' source and target nodes, we can attach the factor $\phi N/2$ to both infection graphs presented in Fig.\ref{fig:infekcje_a}. For dynamic network, the graphs in Fig.\ref{fig:infekcje_b}-\ref{fig:infekcje_c} for infecting a source node through a regular link and through a dynamic link give different counts of how many shortcuts were used. The former was given in Eq.\ref{eq:teor} as $l p$, and the latter actually utilises the same formula, but with the substitution $l\rightarrow l+1$. In total, we get
\beq{}
N_{dyn}=\phi_{dyn} N/2 \cdot l p + \phi_{dyn} N/2 \cdot \left(l+1\right) p \quad.
\eeq

We assume that $N_{dyn}=N_{stat}$ if the epidemic on both networks has the same percolation threshold. Thus, we can obtain the ratio of the two shortcut densities
\beq{anal}
r(p,l)=\phi_{stat}/\phi_{dyn}=\frac{p \left(l+1/2\right)}{T}=\frac{p \left(l+1/2\right)}{1-\left(1-p\right)^l} \quad,
\eeq
where $p$ is the probability of infection in one time step and $l$ latency time of a disease. Now, we can calculate $T_c(r \phi))$ numerically, just as we do it with the fitted $T_c[(1+v)\phi)]$ in Fig.\ref{fig:shift}. The ratio in Eq.\ref{eq:anal} was used to plot the lower solid line in Fig.\ref{fig:shift}.

\section{\label{sec:results} Results}
\subsection{\label{sec:results_shift} Shift of percolation thresholds }
\begin{figure}[htbp]
	\includegraphics[width=\linewidth]{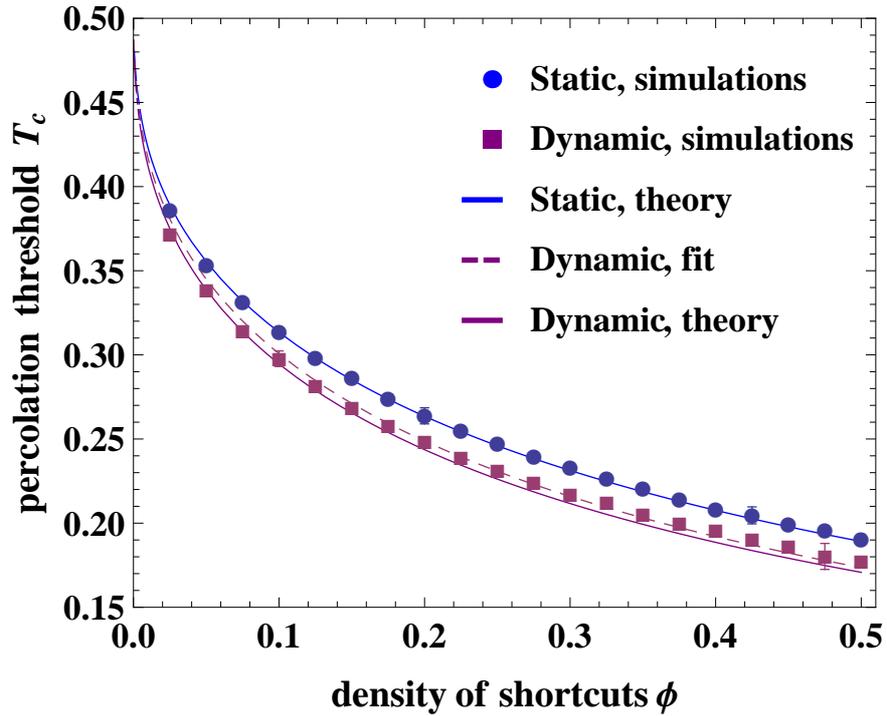}
	\caption{\label{fig:shift} (Colour online) Circles dataset - the static small-world. Squares dataset - the dynamic network. The solid blue (upper) line is the analytic approximation \cite{Newman} for $T_c(\phi)$ and the dashed line gives $T_c[(1+v)\phi)]$, with the fit parameter $v=0.207\pm0.014$. The solid purple (lower) line represents theoretical approximation from Sec.~\ref{sec:dynamics}. Error bars are of the size of the plot markers, unless visible.}
\end{figure}

In Figure \ref{fig:shift} we plot numerical and theoretical values of percolation thresholds $T_c$ for both static and dynamic small-worlds. The resulting $T_c(\phi)$ data points for static small-world network agree with the analytical approximation~\cite{Newman}, which confirms the validity of calibration procedure. As the lower dataset marks the effect of network dynamics, the difference between the two networks proves to be systematic and significant. The dashed line is a fit $T_c[(1+v)\phi]$ of the analytical model for the static network, where the fitted parameter $v$ may be interpreted as a virtual percentage of additional shortcuts needed to obtain the dynamic network percolation thresholds. It follows from the fit that percolation thresholds for dynamic network are lower as if the shortcut density were $(1+v)\phi$ (where $v=0.207\pm0.014$ is the fitted parameter). Nonetheless, qualitatively the epidemic on dynamic small world behaves in the same way as on the static one for the given range of parameters ($\phi=0.5$ corresponds to every node in the network having on average two additional links).

The analytical correction slightly exceeds the values of simulation data points, but the overall agreement is satisfactory. The difference between the analytical solution and the observed behaviour does not exceed the shift between static and dynamic networks obtained from simulations. The discrepancy might be due to the method of calculating percolation thresholds from numerical data or due to the approximate nature of the correction.

\subsection{\label{sec:results_scaling} Suppression of finite-size scaling }

The primary motivation of checking finite-size scaling for the system was to utilise it to determine the percolation thresholds very accurately (as the shift of thresholds observed in Fig.\ref{fig:shift} is relatively small), and to arrive at threshold value for infinite system size. Yet, it is worth noting at this point that the knowledge of thresholds for infinite system sizes would not usually be appropriate for evaluation of risks in the real epidemic, given the sizes of some real networks. To study the size of finite-size effects is thus vital on its own right.

\begin{figure}[p]
\subfigure[]{
\label{fig:scal_fit}
	\includegraphics[width=0.52\linewidth]{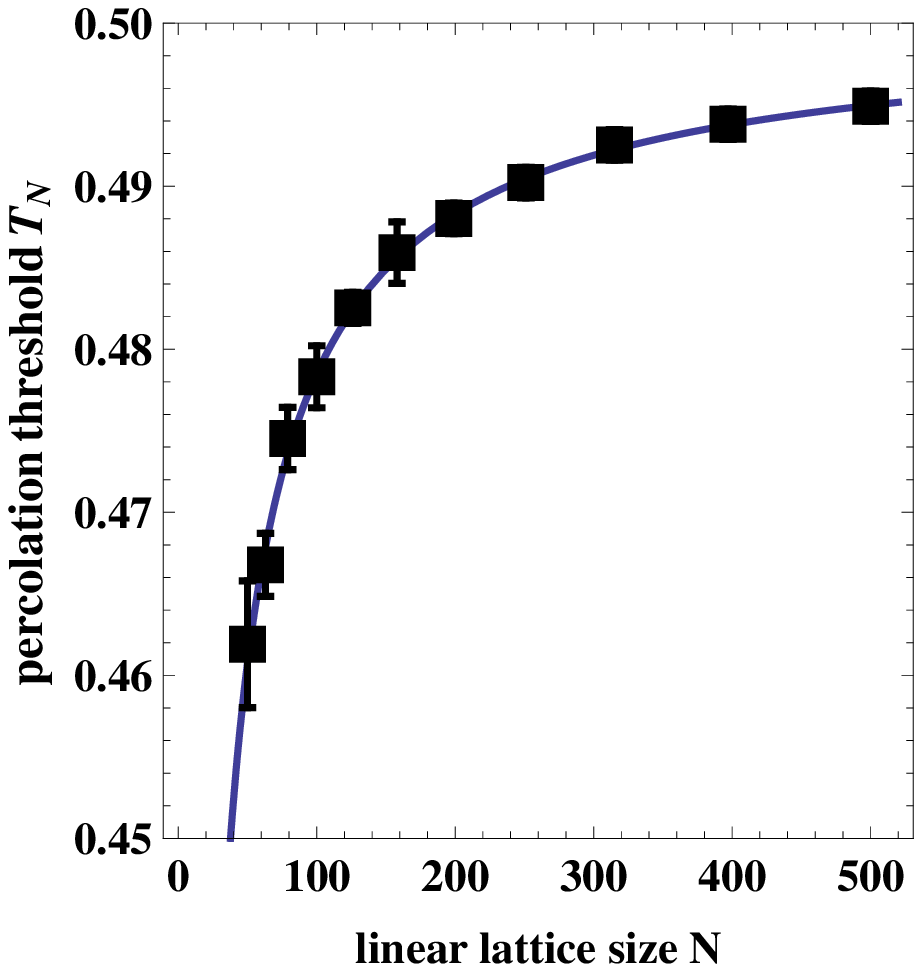}
	}
\subfigure[]{
\label{fig:scal_reg}
	\includegraphics[width=0.47\linewidth]{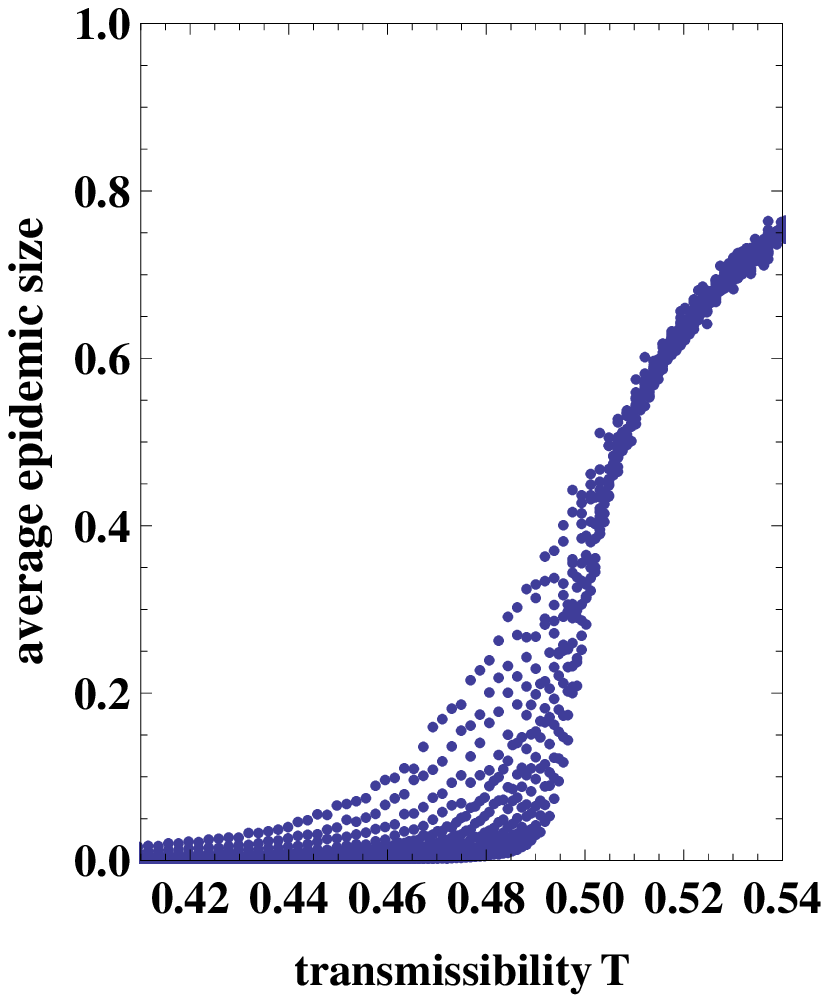}
	}
	\caption{\label{fig:scal} The behaviour of the epidemic outbreak magnitude for various system sizes (linear size vary between $L=50-500$, left- and rightmost data points in \subref{fig:scal_reg}, respectively). \subref{fig:scal_fit} Finite-size scaling $T_N=T_{\infty}-L^{-2/\nu}$ on regular lattice. The points correspond to values of $T$ at the level of the epidemic size $0.1$. \subref{fig:scal_reg} The extent of size dependence for regular lattice.}	
\end{figure}

In Figure \ref{fig:scal_reg} the convergence of the average epidemic size to the threshold behaviour can be observed, and the significant dependence on system size ranges up to the epidemic size of around $0.5 N$ and interval of transmissibility of length around $0.08$ (the numbers are very rough estimates). As presented in Fig.\ref{fig:scal_fit}, one may check that sections of the plot for a given average epidemic size obey scaling of the form
\beq{scaling}
T_N=T_{\infty}-N^{-1/\nu}=T_{\infty}-L^{-2/\nu} \quad,
\eeq
where $T_N$ are the values of transmissibility for a given system size $N$ and a set section position, and $T_{\infty}$ is the percolation threshold for infinite system size. For regular lattice $T_{\infty}$ is fitted correctly for various section positions as $0.500\pm0.005$ (the error may vary for different sections, but does not exceed the given value).

\begin{figure}[p]
\label{fig:scal_dyn}
	\includegraphics[width=0.7\linewidth]{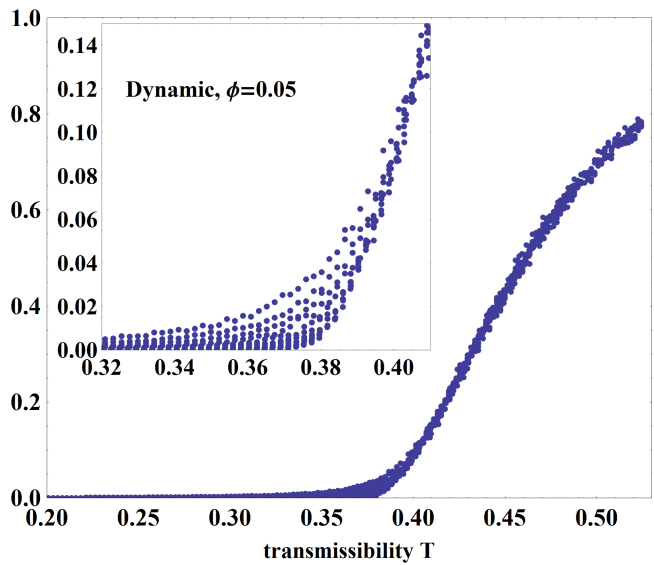}
	\caption{For dynamic small world size dependence of the epidemic outbreak magnitude is suppressed. Inset shows enlarged region around percolation threshold.}	
\end{figure}

It appears that the dependence on system size for small-world networks (both static and dynamic) is dissimilar to the one of regular lattices, as can be seen in Fig.\ref{fig:scal_dyn} ($\phi=0.05$). It is suppressed to smaller values of the average epidemic size. For the shortcuts density $\phi=0.5$ the dependence on system size is already visible only below the epidemic size of $0.03$. Because the dependence of the epidemic size on size of the system becomes of the order of magnitude of statistical fluctuations (the quality of the data can already be seen in the Fig.\ref{fig:scal_dyn}), any attempts to utilise finite-size scaling for determining percolation threshold are not viable. Indeed, the errors do not allow us to check if the same form of finite-size dependence as in Eq.\ref{eq:scaling} holds.


\subsection{\label{sec:dynamics} Dependence on the rate of dynamics}

One can generalise the theoretical analysis for various rates of dynamics, given the formula in Eq.\ref{eq:teor}. To explain this, let us notice that there are two time scales in the model: the latency time $l$ of the infection and the duration $1/d$ between consecutive rewirings of dynamic links (both measured in discrete time steps of the epidemic spread). As the choice of latency $l$ only rescales the total probability of infection $T=T(p,l)$, we can dispose of it, and the crucial parameter $l d$ that accounts for the shift of percolation thresholds is defined as the number of shortcut movements during latency time.

Obviously, for a static network we get $d=0$, while for all the above analysis of dynamic network we have $l d=3$ ($l=3$ and the rewiring was performed every turn, so $d=1$). Depending on the interpretation of the model, we could also consider $d>1$. However, if $p$ is to be the probability of infection \emph{during one time step} it is reasonable that shortcuts rewiring faster than one time step would infect with appropriately smaller probability, and there would be no further shift of percolation thresholds.

Since the epidemic spreads with discrete time, which results in sums as in Eq.\ref{eq:teor}, we are interested in rational numbers $d\in\left[0,1\right]\cap\mathbb{Q}$, particularly of the form $1/i, i\in\mathbb{Z}$. What we need is $N_{dyn}$ calculated in a similar way to that in Eq.\ref{eq:teor}. Here, we take $l=4$, $d=1, 1/2, \ldots, 1/7$, and we plot both the numerical and theoretical results for $\phi=0.25$ in Fig.\ref{fig:dynamics}. Theoretical derivation is to be found in the Appendix. The theoretical approach gives slightly exceeding values (the scale should be noted), which is the same effect as discussed at the end of Section \ref{sec:results_shift}.

\begin{figure}[htbp]
	\includegraphics[width=\linewidth]{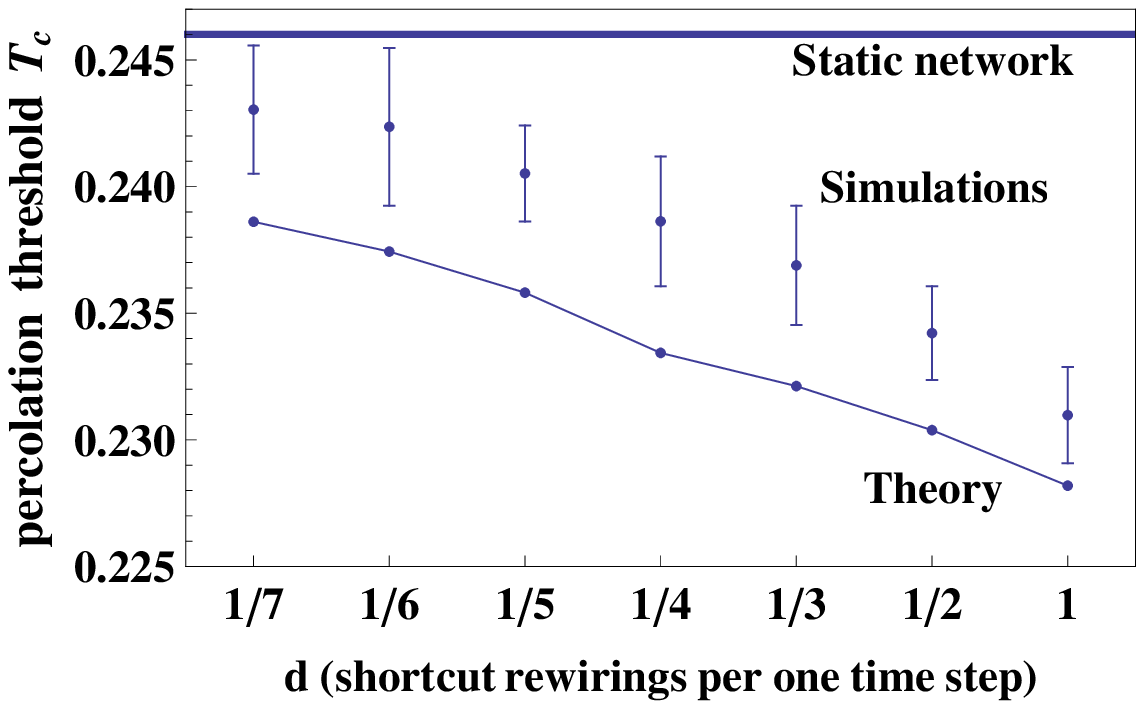}
	\caption{\label{fig:dynamics} Dependence on dynamics for $\phi=0.25$, latency $l=4$.}
\end{figure}

\section{Discussion}

We have shown that introducing dynamics of the long-range links in a small-world network significantly lowers an epidemic threshold in terms of probability of disease transmission, although the overall dependence on number of shortcuts stays the same. Consequently, the risk of an epidemic outbreak is higher than in any calculations involving static models. The effect remains secondary to the influence that the introduction of additional of shortcuts has on the spread of the disease. It should be noted that the shift of percolation thresholds depends on the relative measure of dynamics of the network with respect to the process on the network (rewiring rate and latency time, respectively). Any accurate analytical calculation or simulation should take this quantity as a significant parameter, to be estimated for a particular disease and type of the network.

As in reality we consider only finite-size networks, and real epidemic sizes do not usually reach values of the order of even $10\%$ of the system size, the information on finite-size effects seemed very much needed. That the epidemic outbreak magnitude does not depend on the system size for small-world networks as much as it does for regular lattices means that we should not expect the epidemic outbreaks below transmissibility threshold value. Thus, finite-size effects seem to become secondary, as well.

The usefulness of such a model for risk prediction still depends on our knowledge of the probability of transmission ($p$ or $T$) of a given disease, which is not easy to obtain for diseases spreading outside of well controlled environments like hospitals. Relatively good estimates, thanks to the nature of transmission, exist for syphilis. Transmissibility of the disease is reviewed in \cite{syf}, where authors give values ranging from $9.2\%$ to $63\%$ per partner, and decide on $60\%$ as the lower boundary for \textit{untreated} disease. This seems to be well above the epidemic threshold, irrespective of very different network topology for such diseases. However, this also shows that errors on estimates of transmission probabilities exceed the effect of threshold shifting studied here.

Though the 2-dimensional network structure used here may be said to correspond mainly to that of plantations, it is worth noting its generality: nodes may be interpreted as plants, animals or humans, but also on a larger scale as farms, households, or cities and airports; in turn, long-range links could mean wind (on farms), disease vectors, occasional human contacts, or airline connections.
Still, it has some other fairly realistic characteristics: according to \cite{Eames}, who analysed the structure of human social interactions, `the majority of encounters ($76.70\%;75.26$-$78.07$) occur with individuals never again encountered by the participant during the 14 days of the survey.' This may mean that about $24\%$ of the repeated contacts corresponds roughly to our regular underlying lattice with $z=4$ neighbours for each node, while the $76\%$ correspond to around $3z$ dynamic contacts distributed over over 14 days. This gives on average $\phi\approx0.20$ for simulation with daily time steps, which lies within the parameter range studied in this paper.


\section*{Acknowledgments}
This work is supported by the International PhD Projects Programme of the Foundation for Polish Science within the European Regional Development Fund of the European Union, agreement no. MPD/2009/6.


\section*{\label{sec:app} Appendix: Dependence on the rate of dynamics }

Below we present the way to calculate $N_{dyn}$ for latency periods $l=4,5$ (in the simulation we set $l=4$, but we need to take into account also the process from Fig.\ref{fig:infekcje_c}, which in a sense increases latency by $1$). Let us define

\begin{eqnarray}
		A_{0}(p,l)&=& 1-(1-p)^l \equiv T(l),\quad l\geq1 \nonumber \\
		A_{1}(p,l) 	&=& T(1) \left[1 - T(l-1)\right] + \left[1 - T( 1)\right] T( l-1)+ \nonumber \\
		&+& 2 T(p, 1) T(l-1) \nonumber \\
		A_{2}(p,l) 	&=& T(2) \left[1 - T(l-2)\right] + \left[1 - T(2)\right] T(l-2)+ \nonumber \\
		&+& 2 T(2) T(l-2) \nonumber \\
		A_{11}(p,l)	&=& 3 T(1)^2 T(l-2)+  2 T(1)^2 \left[1 - T(l-2) \right] +\nonumber \\
		&+& 2 T(1) T(l-2) \left[1 - T(1)\right]+ \nonumber \\
		&+& T(l-2)\left[1 - T(1)\right]^2 + \\
		&+&2 T(1) \left[1 - T(1)\right] \left[1 - T(l-2)\right] \nonumber \\
		A_{12}(p,l)	&=& 3 T(1) T(2) T(l-3) + 2 T(1) T(2) \left[1 - T(l-3)\right] +\nonumber\\
		&+& T(1) \left[1 - T(2)\right] T(l-3) +\nonumber\\
		&+& \left[1 - T(1)\right] T(2) T(l-3) \nonumber\\
		&+& T(1) \left[1 - T(2)\right] \left[1 - T(l-3)\right] +\nonumber\\ 
		&+& \left[1 - T(1)\right] T(2) \left[1 - T(l-3)\right] +\nonumber\\
		&+& \left[1 - T(1)\right] \left[1 - T(2)\right] T(l-3), \nonumber
\end{eqnarray}

where we substituted $T(1)$ for $p$ on the right-hand sides, and we leave out the argument $p$ in $T(p,l)$ to simplify the notation. Those quantities correspond to the average number of infections during one latency period depending on when the rewiring takes place. One can present those diagrammatically (here for $l=5$) as

\begin{eqnarray}
		A&_{0}(p,5)&= \cdot \cdot \cdot \cdot \cdot \nonumber \\
		A&_{1}(p,5)&= \cdot | \cdot \cdot \cdot \cdot + \cdot \cdot \cdot \cdot | \cdot = 2 \cdot | \cdot \cdot \cdot \cdot  \nonumber \\
		A&_{2}(p,5)&= \cdot \cdot | \cdot \cdot \cdot + \cdot \cdot \cdot | \cdot \cdot = 2 \cdot \cdot | \cdot \cdot \cdot \label{eq:diagram} \\
		A&_{11}(p,5)&= \cdot | \cdot \cdot \cdot | \cdot \nonumber \\
		A&_{12}(p,5)&= \cdot | \cdot \cdot | \cdot \cdot + \cdot \cdot | \cdot \cdot | \cdot = 2 \cdot | \cdot \cdot | \cdot \cdot \nonumber 
\end{eqnarray}
where the symbol `$|$' marks rewiring, and `$\cdot$' one epidemic time step during latency period. For instance $\cdot | \cdot \cdot$ would correspond to three turns with one rewiring, during which either $0$, $1$ or $2$ infections are possible. The derivation involves only very easy combinatorics, but for longer latency periods one would need to repeat these calculations to obtain more terms and different prefactors.

Now, one can easily obtain expressions for $N_{dyn}$ for any $1/d \in\mathbb{Z}$. Below we give only the general expression for $1/d \geq l$:
\begin{eqnarray}
N_{dyn} &=& \frac{\phi_{dyn} N}{2} d \{\left[2 A_{1}(l) + A_{2}(l) + (1/d + 1 - l) A_{0}(l)\right]+\nonumber \\
&+&\left[2 A_{1}(l + 1) + 2 A_{2}(l + 1) + (1/d - l) A_{0}(l)\right]\}
\end{eqnarray}
where $l=4$. The first term in the brackets corresponds to Fig.\ref{fig:infekcje_b} and the second to Fig.\ref{fig:infekcje_c}. For greater numbers of rewiring per turn $d$, we need to consider the terms $A_{11}, A_{12}$. The result is plotted against simulated data in Fig.\ref{fig:dynamics}.


\end{document}